\documentclass[11pt,a4paper,useAMS]{emulateapj}

\usepackage{natbib}
\usepackage{amsmath}
\usepackage{graphicx}
\usepackage{morefloats}

\begin{document}


\title{Complex diffuse radio emission in the merging PLANCK ESZ cluster Abell 3411}


\author{
R.~J.~van~Weeren\altaffilmark{1},  K.~Fogarty\altaffilmark{1,2}, C.~Jones\altaffilmark{1},  W.~R.~Forman\altaffilmark{1}, T.~E.~Clarke\altaffilmark{3}, M.~Br\"uggen\altaffilmark{4},  
R.~P.~Kraft \altaffilmark{1}, D.~V.~Lal\altaffilmark{5}, 
S.~S.~Murray\altaffilmark{1,6}, and H.~J.~A.~R\"ottgering\altaffilmark{7} \vspace{3mm}
}

\affil{\altaffilmark{}}

\affil{\altaffilmark{1}Harvard-Smithsonian Center for Astrophysics, 60 Garden Street, Cambridge, MA 02138, USA}

\affil{\altaffilmark{2} Department of Physics and Astronomy, The Johns Hopkins University, 3400 N. Charles St., Baltimore, MD 21218-2686, USA}

\affil{\altaffilmark{3}Naval Research Laboratory Remote Sensing Division, Code 7213 4555 Overlook Ave SW, Washington, DC 20375, USA}
\affil{\altaffilmark{4}Hamburger Sternwarte, Gojenbergsweg 112, 21029 Hamburg, Germany}
\affil{\altaffilmark{5}National Centre for Radio Astrophysics, TIFR, Pune University Campus, Post Bag 3, Pune 411 007, India}
\affil{\altaffilmark{6}Department of Physics and Astronomy, Johns Hopkins University, 3400 North Charles Street, Baltimore, MD 21218, USA}
\affil{\altaffilmark{7}Leiden Observatory, Leiden University, P.O. Box 9513, NL-2300 RA Leiden, The Netherlands}

\email{E-mail: rvanweeren@cfa.harvard.edu}

\altaffiltext{4}{Einstein Fellow}

\shorttitle{Diffuse radio emission in Abell~3411}
\shortauthors{VAN WEEREN ET AL.}

\vspace{0.5cm}
\begin{abstract}
\noindent  We present \textit{VLA} radio and \textit{Chandra} X-ray observations of the merging galaxy cluster \object{Abell~3411}. For the cluster, we find an overall temperature of 6.4$^{+0.6}_{-1.0}$~keV and an X-ray luminosity of  $2.8 \pm 0.1 \times 10^{44}$~erg~s$^{-1}$ between 0.5 and 2.0~keV. The \textit{Chandra} observation reveals the cluster to be undergoing a merger event. The \textit{VLA}  observations show the presence of large-scale diffuse emission in the central region of the cluster, which we classify as a 0.9~Mpc size radio halo. In addition, a complex region of diffuse, polarized emission is found in the southeastern outskirts of the cluster, along the projected merger axis of the system. 
We classify this region of diffuse emission as a radio relic. 
The total extent of this radio relic is 1.9~Mpc. For the combined emission in the cluster region, we find a radio spectral index of $-1.0 \pm0.1$ between 74~MHz and 1.4~GHz.
The morphology of the radio relic is peculiar, as the relic is broken up into five fragments. This suggests that the shock responsible for the relic has been broken up due to interaction with a large-scale galaxy filament connected to the cluster or other substructures in the ICM. Alternatively, the complex morphology reflects the presence of electrons in fossil radio bubbles that are re-accelerated by a shock.  
\end{abstract}





\section{Introduction}

In a number of clusters, large-scale radio emission is found in the form of radio halos and relics \citep[e.g.,][and references therein]{2012A&ARv..20...54F}. This large-scale diffuse emission indicates the presence of relativistic particles, with Lorentz factors  $\gamma \sim 10^{4}$,  and  $\mu$Gauss magnetic fields in the intracluster medium (ICM). 
Giant radio halos and relics, i.e., those having Mpc-sizes, are exclusively found in merging galaxy clusters \citep[e.g.,][]{2010ApJ...721L..82C,2011A&A...533A..35V}.  This supports the picture that a fraction of the energy released during the merger event is channeled into the (re-)acceleration of particles to highly relativistic energies. In the presence of a magnetic field these particles would then emit faint synchrotron radiation at radio wavelengths. 

{Giant radio halos} have $\sim 1$--2~Mpc sizes, are centrally located in clusters,  display smooth morphologies, and are usually unpolarized \citep[e.g.,][]{2001A&A...373..106F, 2003A&A...400..465B}. Creating these large halos requires that local particle acceleration occurs over a large volume in the cluster \citep{1977ApJ...212....1J}. Although the basic observational properties of halos have now been established  \citep[e.g.,][]{2012A&ARv..20...54F}, the formation mechanism of halos is still being debated \citep[e.g.,][]{2008Natur.455..944B, 2010MNRAS.401...47D, 2010MNRAS.407.1565D, 2010A&A...517A..43M, 2011MNRAS.412....2B,  2011A&A...527A..99E, 2012MNRAS.426..956B, 2012arXiv1207.6410Z,2012ApJ...757..123A}.
Scenarios for their origin include ``primary'' models, in which an existing electron population is re-accelerated by turbulence or shocks caused by recent cluster mergers \citep[][]{2001MNRAS.320..365B, 2001ApJ...557..560P}, and ``secondary'' models, in which relativistic electrons are continuously injected into the ICM by inelastic collisions between cosmic rays and thermal ions \citep[e.g.,][]{1980ApJ...239L..93D, 1999APh....12..169B, 2000A&A...362..151D, 2001ApJ...562..233M, 2010ApJ...722..737K}.  Combinations of both acceleration mechanisms have also been considered  \citep{2005MNRAS.363.1173B,2008SSRv..134..311D,  2011MNRAS.410..127B}.

Giant {radio relics} are located in cluster outskirts, are usually polarized, and are elongated \citep[e.g.,][]{1997MNRAS.290..577R, 2011ApJ...736L...8B}. They sometimes occur in pairs, located diametrically on opposite sides of the cluster center, so-called ``double relics''. The currently favored scenario is that large relics are the signatures of electrons (re-)accelerated at shocks due to cluster mergers \citep[e.g.,][]{1998A&A...332..395E,2001ApJ...562..233M, 2008A&A...486..347G, 2009A&A...494..429B,2010Sci...330..347V,2010ApJ...718..939K}. This should happen via the diffusive shock acceleration mechanism \citep[DSA; e.g.,][]{1977DoSSR.234R1306K,1977ICRC...11..132A,1978MNRAS.182..147B,1978MNRAS.182..443B,1978ApJ...221L..29B, 1983RPPh...46..973D, 1987PhR...154....1B,1991SSRv...58..259J, 2001RPPh...64..429M,2013ApJ...764...95K}. Cluster merger shocks typically have low Mach numbers \citep[$\mathcal{M} \sim 1-3$, e.g.,][]{2002ApJ...567L..27M,2010MNRAS.406.1721R,2011ApJ...728...82M,2013PASJ...65...16A,2013MNRAS.429.2617O} and the efficiency with which such shocks can accelerate particles is unknown. Therefore, re-acceleration of pre-accelerated electrons in the ICM might be required to explain the observed brightness of relics, since re-acceleration is a more efficient mechanism for weak shocks \citep[e.g.,][]{2005ApJ...627..733M,2008A&A...486..347G,2011ApJ...734...18K,2012ApJ...756...97K, 2013arXiv1301.5644P}.

Only a few dozen clusters with diffuse radio emission are known today 
 \citep{2007A&A...463..937V, 2008A&A...484..327V, 2009A&A...507.1257G, 2011A&A...533A..35V, 2012MNRAS.420.2006N,2012MNRAS.426...40B}. 
This makes it difficult to study the halo and relic occurrence as a function of cluster properties, such as the dynamical state, in detail.

To identify both a larger number of clusters with radio halos and relics and a sample with more extreme clusters in terms of their dynamics and structural morphologies, we compared the radio emission as mapped by the NVSS \citep{1998AJ....115.1693C} and \textit{Very Large Array (VLA)} with \textit{Chandra} X-ray images of the Planck ESZ clusters \citep{2011A&A...536A...8P}.  Sunyaev-Zel'dovich surveys preferentially select massive clusters, including those in early stages of formation  \citep{2011A&A...536A...9P,2012A&A...543A.102P}.

Here we report on \textit{VLA}, \textit{Chandra} and Very Large Telescope (VLT) observations of the poorly studied cluster Abell~3411 located at $z=0.1687$ \citep{2002ApJ...580..774E}. It is listed in the Planck ESZ catalog as
PLCKESZ G241.97+14.85 \citep{2011A&A...536A...8P}. Observations and data reduction are described in Sect.~\ref{sec:obs}. The results are presented in Sect.~\ref{sec:results} and this is followed by a discussion and conclusions in Sects. \ref{sec:discussion} and \ref{sec:conclusions}.

{Throughout this paper we assume a $\Lambda$CDM cosmology with $H_{0} = 71$~km~s$^{-1}$~Mpc$^{-1}$, $\Omega_{m} = 0.27$, and $\Omega_{\Lambda} = 0.73$. With the adopted cosmology, 1\arcsec~corresponds to a physical scale of 2.85~kpc at $z=0.1687$}. All images are in the J2000 coordinate system.

\begin{figure*}
\begin{center}
\includegraphics[angle =90, trim =0cm 0cm 0cm 0cm,width=0.49\textwidth]{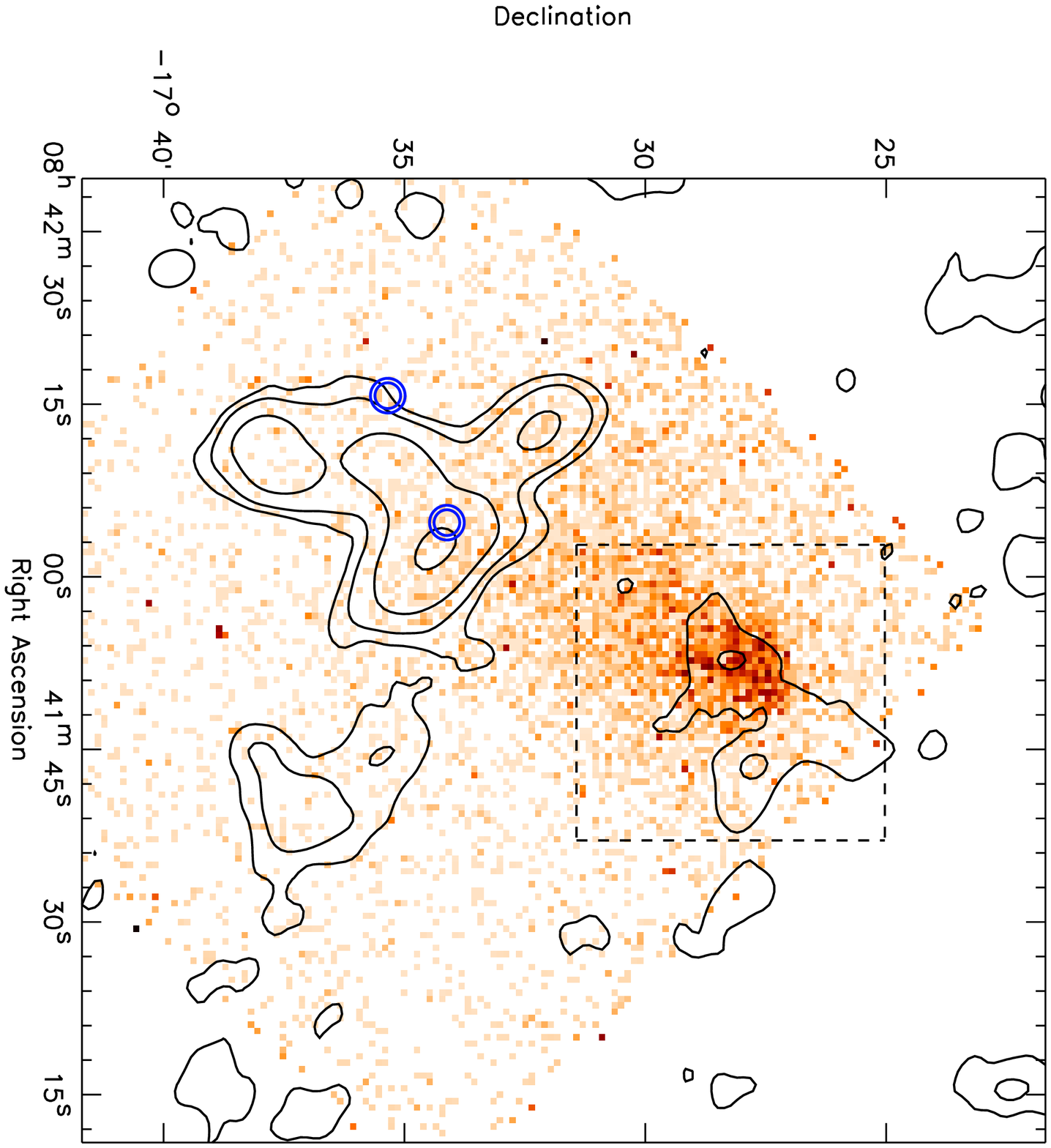}
\includegraphics[angle =90, trim =0cm 0cm 0cm 0cm,width=0.49\textwidth]{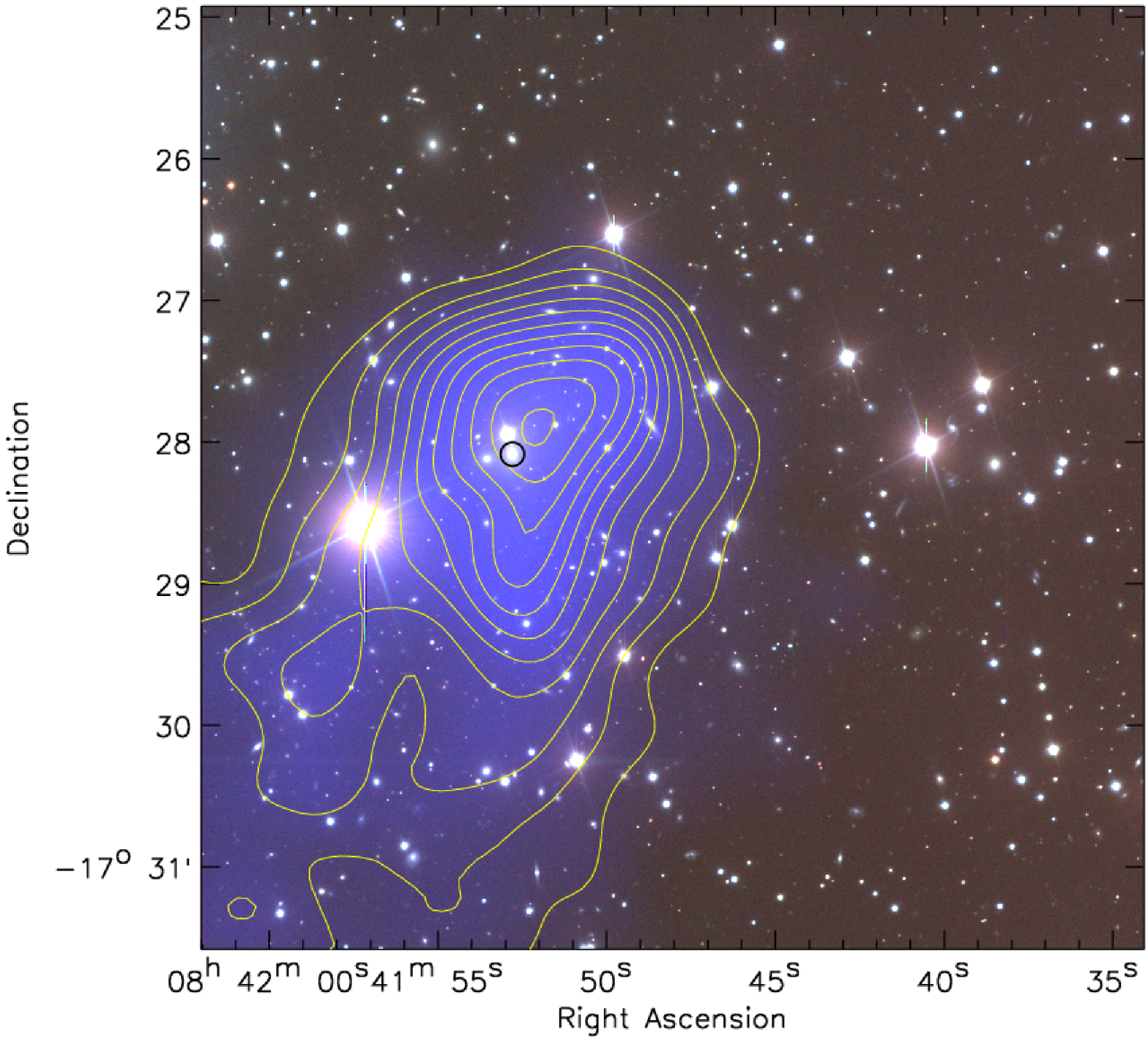}
\end{center}
\caption{Left: \textit{Chandra} ACIS~0.5--3.0~keV image of Abell~3411. Radio contours are from the \textit{VLA} 1.4~GHz DnC array image with natural weighting and compact sources subtracted. The resolution of the radio image is $57\arcsec \times 45\arcsec$.  Contour levels are drawn at $[1, 3, 9, 27, \ldots] \times 3\sigma_{\mathrm{rms}}$, with $\sigma_{\mathrm{rms}}=55$~$\mu$Jy~beam$^{-1}$. The locations of two compact sources, labeled in Figs.~\ref{fig:vlaimages} and \ref{fig:optical}, are indicated with blue circles. The dashed box indicates the region shown in the right panel. Right: VLT FORS1 color image (using V$_{\rm{Bessel}}$, R$_{\rm{Bessel}}$, and I$_{\rm{Bessel}}$ filters) of the central part of the cluster. \textit{Chandra} ACIS~0.5--3.0~keV adaptively smoothed image is overlaid in blue, yellow contours are linearly spaced for the X-ray surface brightness. The cD galaxy 2MASX~J08415287-1728046 is marked with a black circle.}
\label{fig:xray}
\end{figure*}

\section{Observations \& data reduction}

\label{sec:obs}

Abell 3411 was observed with the \textit{VLA} on October 7, 2003 in BnA array and on June 14, 2004 in DnC array (project AC0696). 
An overview of the observations and resulting images is given in Table \ref{tab:observations}. The 1.4~GHz observations were carried out in single channel continuum mode with two IFs with a bandwidth of 50 MHz. All polarization products (RR, RL, LR, and LL) were recorded. 

The data were calibrated with the NRAO Astronomical Image Processing System (AIPS) package. The data were visually inspected for the presence of radio frequency interference, which was subsequently flagged. We then determined the gains for the calibrator sources. The fluxes for the primary calibrator sources (3C147 and 3C286) were set by the \cite{perleyandtaylor} extension to the \cite{1977A&A....61...99B} scale. 
The effective feed polarization parameters (the leakage terms or D-terms) were found by observing the phase calibrator (0902-142) over a range of parallactic angles and simultaneously solving for the unknown polarization properties of the source. 
The polarization angles were set using the polarized source 3C~286.  For the R-L phase difference of 3C~286, we assumed a value of $-66.0\degr$. 
We transferred the calibrator solutions to the target and carried out several rounds of self-calibration to refine the calibration. 
Images were made with manually placed clean boxes and the ``briggs'' weighting scheme \citep[][see Table~\ref{tab:observations}]{briggs_phd}, unless explicitly stated. {All final images were corrected for the primary beam attenuation, unless noted.}

\begin{table}
\begin{center}
\caption{VLA Observations}
\begin{tabular}{lllll}
\hline
\hline
& VLA BnA & VLA DnC & \\
\hline
Frequency (VLA band)                & 1365, 1435~MHz & 1365, 1435~MHz\\
Bandwidth 		  & $ 2\times 50$~MHz & $2\times 50$~MHz  \\
Observation dates				& 7 Oct, 2003&  14 Jun, 2004   \\
Total on-source time (hr)		&  4.2 & 4.6\\
Beam size			& $7.0\arcsec \times 3.9\arcsec$ & $48\arcsec \times 33\arcsec$ \\
rms noise$^a$ ($\sigma_{\rm{rms}}$)	& 15 $\mu$Jy beam$^{-1}$ &   47  $\mu$Jy beam$^{-1}$\\	
Robust  (briggs weighting)   & $-0.5$ &1.0   \\
\hline
\hline
\end{tabular}
\label{tab:observations}
\\
{$^a$ the rms noise is measured at the center of the images, between the radio halo and relic (see Figs.~\ref{fig:xray} and \ref{fig:vlaimages})}
\end{center}
\end{table}

The cluster was also observed using the {\textit Chandra X-ray Observatory} for 10 ks using the ACIS-I array. The data (\dataset [ADS/Sa.CXO#obs/13378] {Chandra ObsId 13378}) were calibrated by applying the most recent calibration files to the level 1 events file. The events file was calibrated following the processing described in \cite{2005ApJ...628..655V}. This included correction for position dependent charge transfer inefficiency, the application of gain and time-dependent gain maps to calibrate photon energies, and removal of counts from bad pixels and counts with a recomputed ASCA grade 1, 5, or~7. Periods of high local background were removed from the observation by examining the light curve and removing periods with a flux in the 5--10~keV band 1.2~times above the mean. This left a total exposure of 7.0~ks.
For the image in Fig.~\ref{fig:xray} (left panel), we extracted counts in the 0.5--3.0~keV band to maximize the ratio between source and background counts in the image. Pixels in the image were combined with a binning factor of~16.

VLT FORS1 images of Abell~3411 were obtained from the ESO archive. These images were taken on 31 December, 2003, with V$_{\rm{Bessel}}$, R$_{\rm{Bessel}}$, and I$_{\rm{Bessel}}$ filters and only cover the central region of the cluster. In total 3 images with an exposure time of 110~s each were available per filter. These images were bias-corrected, flat-fielded, and stacked using IRAF  \citep{1986SPIE..627..733T, 1993ASPC...52..173T}.

\section{Results}
\label{sec:results}

\begin{figure*}
\begin{center}
\includegraphics[angle =90, trim =0cm 0cm 0cm 0cm,width=0.49\textwidth]{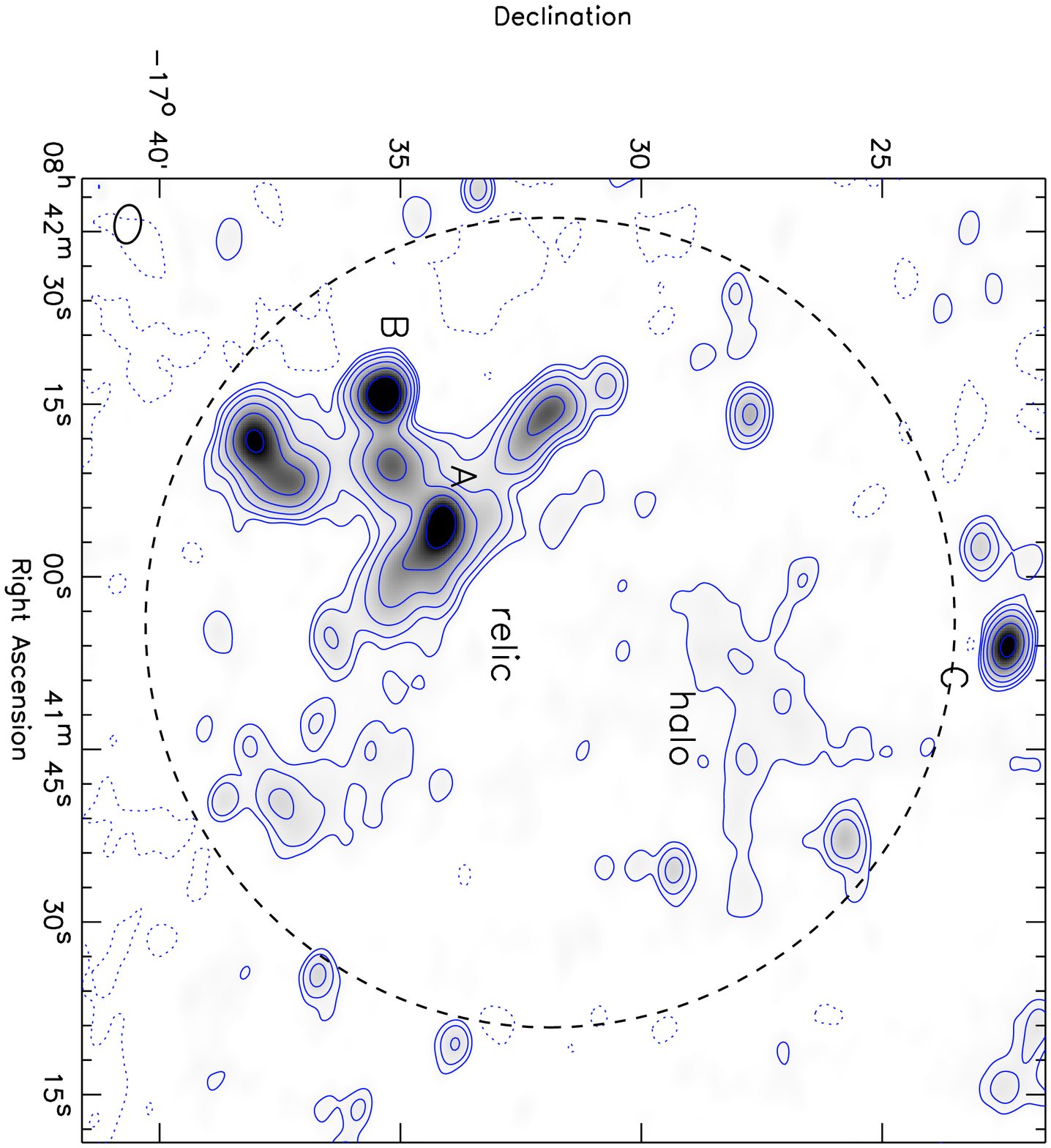}
\includegraphics[angle =90, trim =0cm 0cm 0cm 0cm,width=0.49\textwidth]{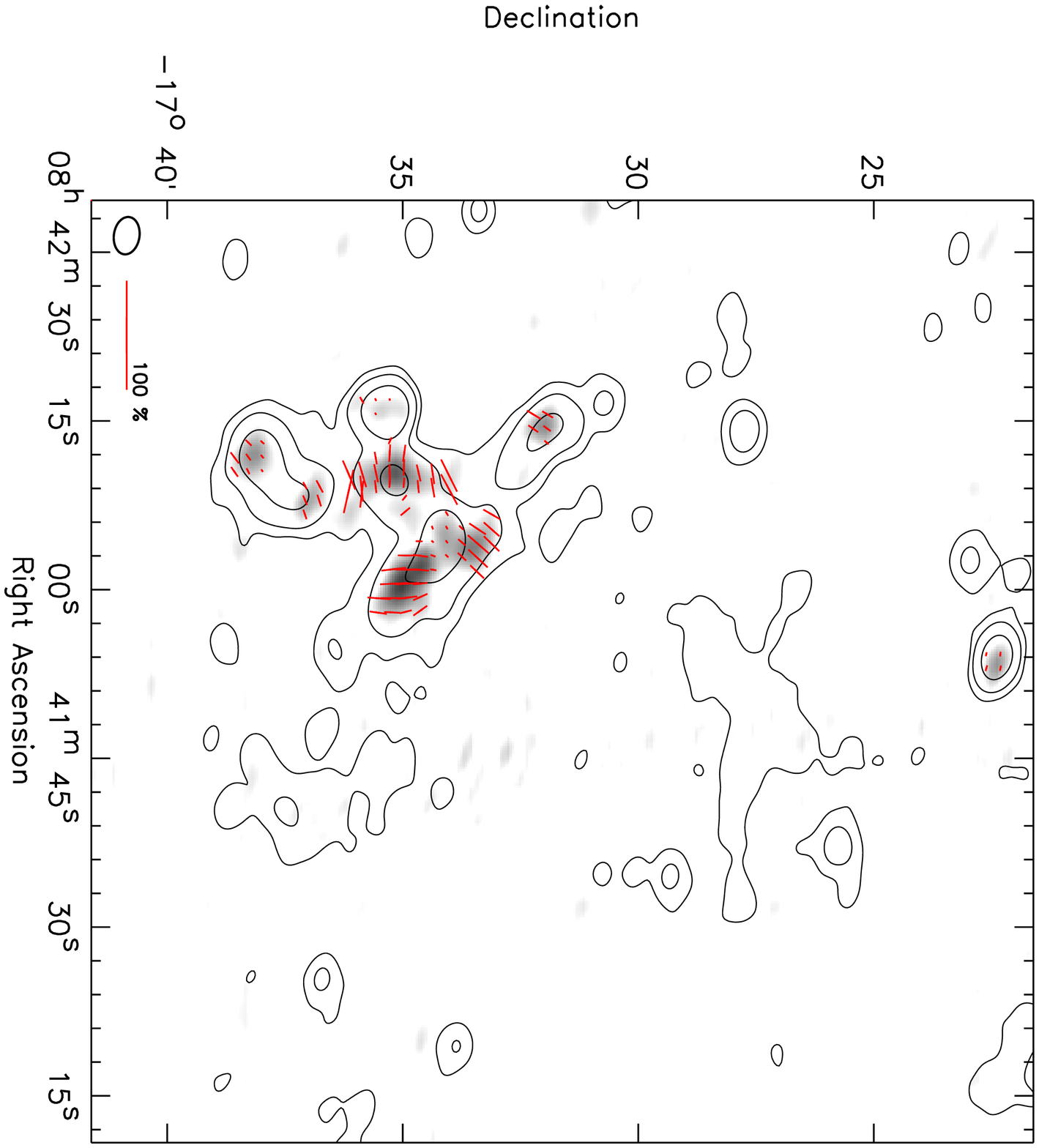}
\end{center}
\caption{{Left: \textit{VLA} 1.4~GHz DnC array image. Contour levels are drawn at $[1, 2, 4, 8, \ldots] \times 3\sigma_{\mathrm{rms}}$ ($\sigma_{\mathrm{rms}}= 47$~$\mu$Jy~beam$^{-1}$). Negative  $-3\sigma_{\mathrm{rms}}$ contours are shown by the dotted lines. {The dashed circle indicates the area where the flux densities were extracted for Fig.~\ref{fig:flux}}. Right: \textit{VLA} DnC array polarization map. The image is not corrected for the primary beam attenuation as the polarization vectors and polarization fraction are independent of the beam correction. Total polarized intensity 
is shown as a grayscale image.  Red vectors depict the polarization E-vectors, 
their length represents the polarization fraction. A reference vector for a polarization fraction of 100\% is shown in 
the bottom left corner. No vectors were drawn for pixels with a 
SNR $< 4$ in the total polarized intensity image. 
Contour levels are drawn at ${[1, 4, 16, 64, \ldots]} \times 3\sigma_{\mathrm{rms}}$  ($\sigma_{\mathrm{rms}}= 47$~$\mu$Jy~beam$^{-1}$)
and are from the Stokes I image.}}
\label{fig:vlaimages}
\end{figure*}

\begin{figure*}
\begin{center}
\includegraphics[angle =90, trim =0cm 0cm 0cm 0cm,width=0.49\textwidth]{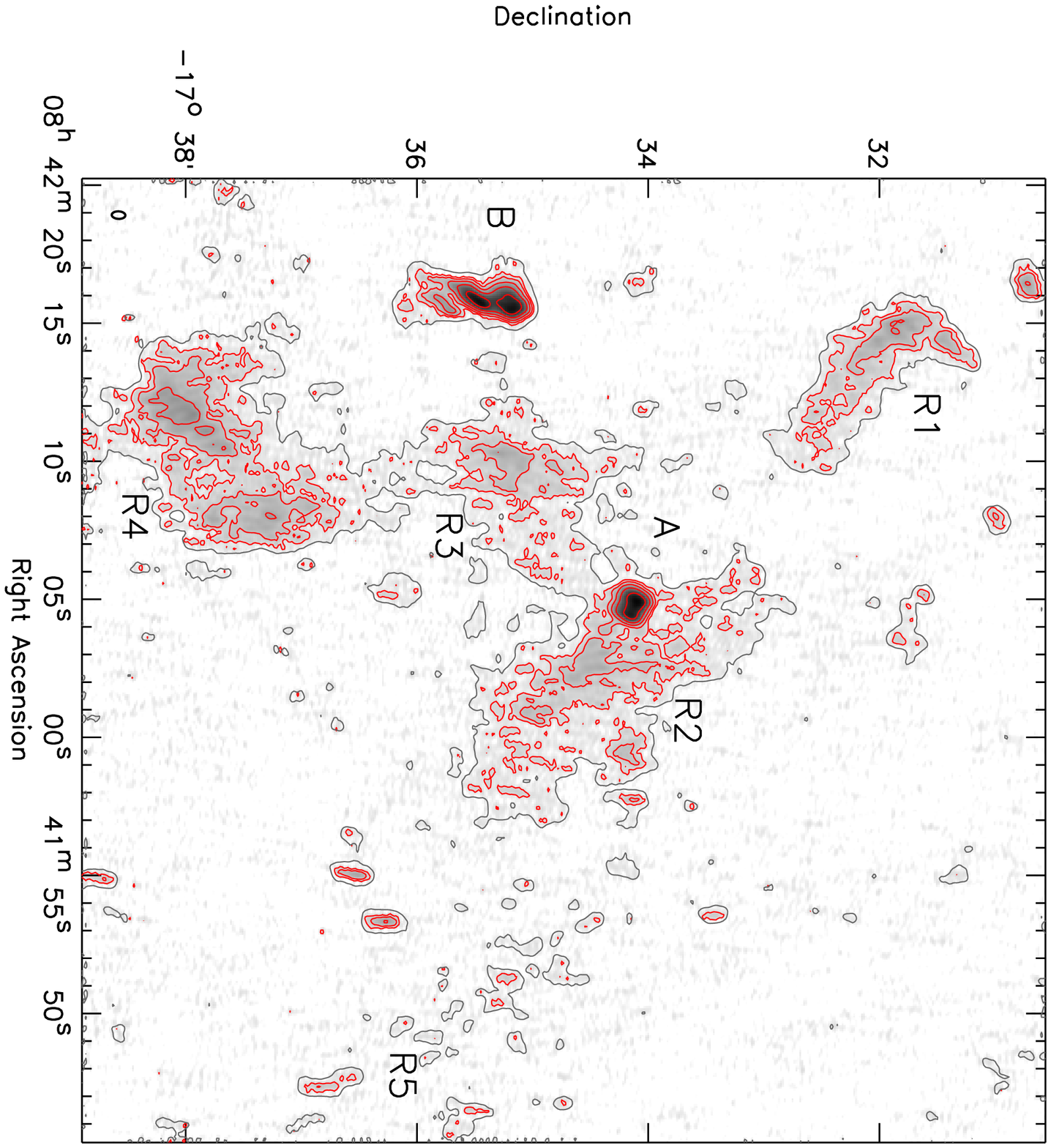}
\includegraphics[angle =90, trim =0cm 0cm 0cm 0cm,width=0.49\textwidth]{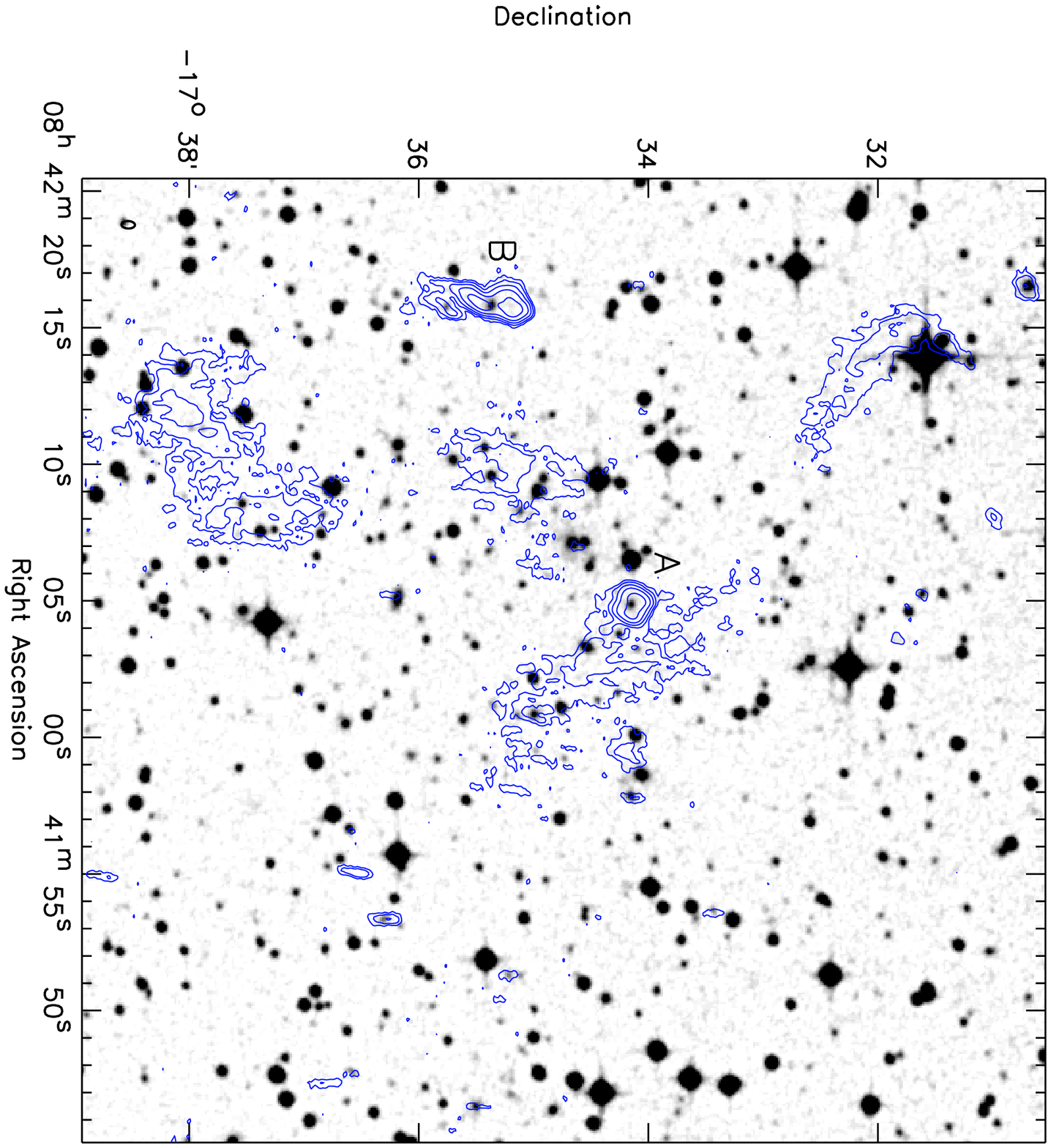}
\end{center}
\caption{Left:  \textit{VLA} 1.4~GHz BnA array image of the relic. Red contour levels are drawn at $[1, 2, 4, 8, \ldots] \times 4\sigma_{\mathrm{rms}}$ ($\sigma_{\mathrm{rms}}= 15$~$\mu$Jy~beam$^{-1}$). Grey contours are drawn at $1.5\sigma_{\mathrm{rms}}$ using a boxcar average of 7\arcsec. The beam size is shown at the bottom left corner of the images. Right: Optical DSS2 red image around sources $A$ and $B$ overlaid with  radio contours from the \textit{VLA}~BnA array image. Contours levels are drawn at $[1, 2, 4, 8, \ldots] \times 4\sigma_{\mathrm{rms}}$. 
}
\label{fig:optical}
\end{figure*}

The \textit{Chandra} ACIS observations of Abell 3411 show a merging system, with the projected merger axis oriented SE-NW (see Fig.~\ref{fig:xray}). The cluster has a cometary shape with a well-defined subcluster core visible in the northwestern part of the system. Fainter X-ray emission is found surrounding the subcluster core and this emission seems to be part of a second larger subcluster that likely extends beyond the northern boundary of the Chandra image.
{There is no clear surface brightness peak corresponding to the primary cluster core, which suggests the primary cluster has been disrupted by the collision with the subcluster, as has been the case for \object{Abell~2146} \citep{2012MNRAS.423..236R}.} The ICM temperature was fit using {\tt XSPEC} with an absorbed single-temperature APEC model, which describes emission from collisionally-ionized diffuse plasma. The hydrogen column density was fixed using \cite{1990ARA&A..28..215D} {at $5.39 \times 10^{20}$~cm$^{-2}$}. The temperatures and metallicities were left free in the fit. From this we found a global temperature of $6.4^{+0.6}_{-1.0}$~keV. Since $R_{500}$ partially extends beyond the Chandra field of view (FOV) we calculated the $R_{500}$-luminosity using a surface brightness profile. The (unabsorbed) X-ray luminosity is found to be $2.8 \pm 0.1 \times 10^{44}$~erg~s$^{-1}$ between 0.5 and 2.0~keV within $R_{500}=1.34$~Mpc. 
This would correspond to a mass of about $M_{500}=9 \times 10^{14}$~$M_{\odot}$ \citep{2009ApJ...692.1033V}.  

The VLT FORS1 image is shown in Fig.~\ref{fig:xray} right panel. A cD galaxy is visible just southeast of the X-ray peak and at about 10\arcsec~south of a foreground star (the cD galaxy is marked in Fig.~\ref{fig:xray} right panel). This galaxy, \object{2MASX J08415287-1728046} has a measured redshift of $0.16324$  \citep{2009MNRAS.399..683J}, consistent with the cluster redshift. The image displays a relatively large number of foreground stars as A3411 is located at a galactic latitude of $b\approx15\degr$.

\begin{table}
\begin{center}
\caption{Cluster Properties}
\begin{tabular}{lllll}
\hline
\hline
Redshift                         & 0.1687 \\
$L_{\rm{X, R_{500}}}$   (erg~s$^{-1}$, 0.5--2.0~keV$^{a}$)              & $2.8 \pm 0.1 \times 10^{44}$ \\
Temperature  (keV)		  &  6.4$^{+0.6}_{-1.0}$  \\
LLS relic, halo (Mpc)   &  1.9, 0.9   \\
$P_{\rm{1.4 GHz}}$ relic, halo (W~Hz$^{-1}$)		&  $5.0 \times 10^{24}$, $4.6 \times 10^{23}$ \\
$R_{\rm{projected}}$ relic  (Mpc) & 1.3 \\
\hline
\hline
\end{tabular}
\label{tab:cluster}
\end{center}
$^{a}$ \cite{2002ApJ...580..774E} reported  $L_{\rm{X}}=3.8 \times 10^{44}$ from ROSAT in the 0.1--2.4~keV band (corrected for our adopted $\Lambda$CDM cosmology)\\
\end{table}

The 1.4~GHz \textit{VLA} continuum images are shown in Figs.~\ref{fig:vlaimages} and \ref{fig:optical}. A low-resolution image, with the emission from compact sources subtracted, is shown as an overlay on the \textit{Chandra} image in Fig.~\ref{fig:xray}. For this, we first created an image of the compact sources using uniform weighing and an inner uv-range cut of 2~k$\lambda$. We then subtracted the clean components of this image from the uv-data and re-made the radio image using natural weighting.

Two relatively bright compact sources are visible to the SE of the cluster center in the higher resolution BnA array image. These sources are labeled $A$ and $B$ in Figs.~\ref{fig:vlaimages} and \ref{fig:optical}. {We measure flux densities of $9.5\pm0.5$ and $16.1\pm 0.8$ in the BnA array image for source $A$ and $B$, respectively.}
Both sources have optical counterparts in the Digital Sky Survey (DSS) image (see Fig.~\ref{fig:optical} right panel for an overlay of radio contours on a DSS image). No redshifts are available for these counterparts. With the 2MASS Ks band magnitudes \citep[mag$_{\rm{Ks}}=$15.1 for both $A$ and $B$, ][]{2006AJ....131.1163S} we estimate redshifts of $\sim0.3$ for both galaxies using the K-z relation for massive elliptical galaxies from \cite{2003MNRAS.339..173W}. The uncertainty in this estimate is about $0.1$ so either these galaxies are cluster members or they are located behind the cluster. Source $A$ is resolved into two components that are separated by about 8\arcsec. The morphology of source $B$ is more complex and it has a largest angular extent of 53\arcsec. At the distance of A3411  53\arcsec~corresponds to a physical extent of about 150~kpc. Merging galaxy clusters often host disturbed and/or tailed radio sources \citep[e.g.,][]{1979A&A....80..201B} so it seems likely that at least source $B$ is associated with A3411. 

The only other galaxy with a published redshift in the A3411 field, besides the central cD galaxy, is \object{2MASX J08415393-1722206}. This galaxy is associated with the bright compact radio source $C$ at the far north in Fig.~\ref{fig:vlaimages} (left panel) and is likely a cluster member with $z=0.162569$ \citep{2009MNRAS.399..683J}. {We measure a flux density of $9.1 \pm 0.5$~mJy for the source in the BnA array image.}

The radio images display a region of complex radio emission to the SE of the cluster, along the major axis of the extended X-ray emission. The diffuse emission to the SE consists of five elongated components (labelled R1 to R5, Fig.~\ref{fig:optical} left), with a total extent of 11\arcmin in the NE-SW direction, corresponding to 1.9~Mpc at the distance of A3411. Four of those components can be seen in the higher-resolution BnA array image (Fig.~\ref{fig:optical} left). The fifth fainter component (R5) extents further to the SW (see Fig.~\ref{fig:xray} left). The source (the part around $A$) is located at a projected distance ($R_{\rm{projected}}$) of 1.3 Mpc from the X-ray peak of the cluster. We measure an integrated flux of $53\pm3$~mJy for the combined emission from R1 to R5 in the image with the compact sources subtracted (Fig.~\ref{fig:xray} left panel). This corresponds to a radio power of $P_{\rm{1.4~GHz}} = 5.0 \times 10^{24}$~W~Hz$^{-1}$. In addition, we detect polarized emission from this diffuse source at the 10 to 25\% level (see Fig.~\ref{fig:vlaimages} right). The emission cannot be classified as the lobes of an FR-I type radio source \citep{1974MNRAS.167P..31F} related to either source $A$ or $B$, since the counterparts of $A$ and $B$ are located in or behind the cluster, making the source too large for an FR-I type radio galaxy. Instead, we classify this complex radio source as a radio relic due to its size, location with respect to cluster center, lack of optical counterparts, and polarization properties. 

In the central part of the cluster diffuse and faint extended radio emission is found which we classify as a radio halo.   The halo appears to be elongated along the major axis of the ICM and has a physical extent of $\sim0.9$~Mpc. 
We measure a total flux in the halo region of $4.8\pm0.5$~mJy which corresponds to $P_{\rm{1.4~GHz}} = 4.6 \times 10^{23}$~W~Hz$^{-1}$. Most of the halo is only detected at $\sim3\sigma_{\rm{rms}}$. The actual halo power might be somewhat higher as the total extent of the halo could be larger. The radio power we find is consistent with the X-ray luminosity-radio power ($L_{X}-P_{1.4}$) correlation for radio-halo clusters \citep[e.g.,][]{2000ApJ...544..686L,2002A&A...396...83E,2006MNRAS.369.1577C}. 
\begin{figure*}
\begin{center}
\includegraphics[angle =90, trim =0cm 0cm 0cm 0cm,width=0.4875\textwidth]{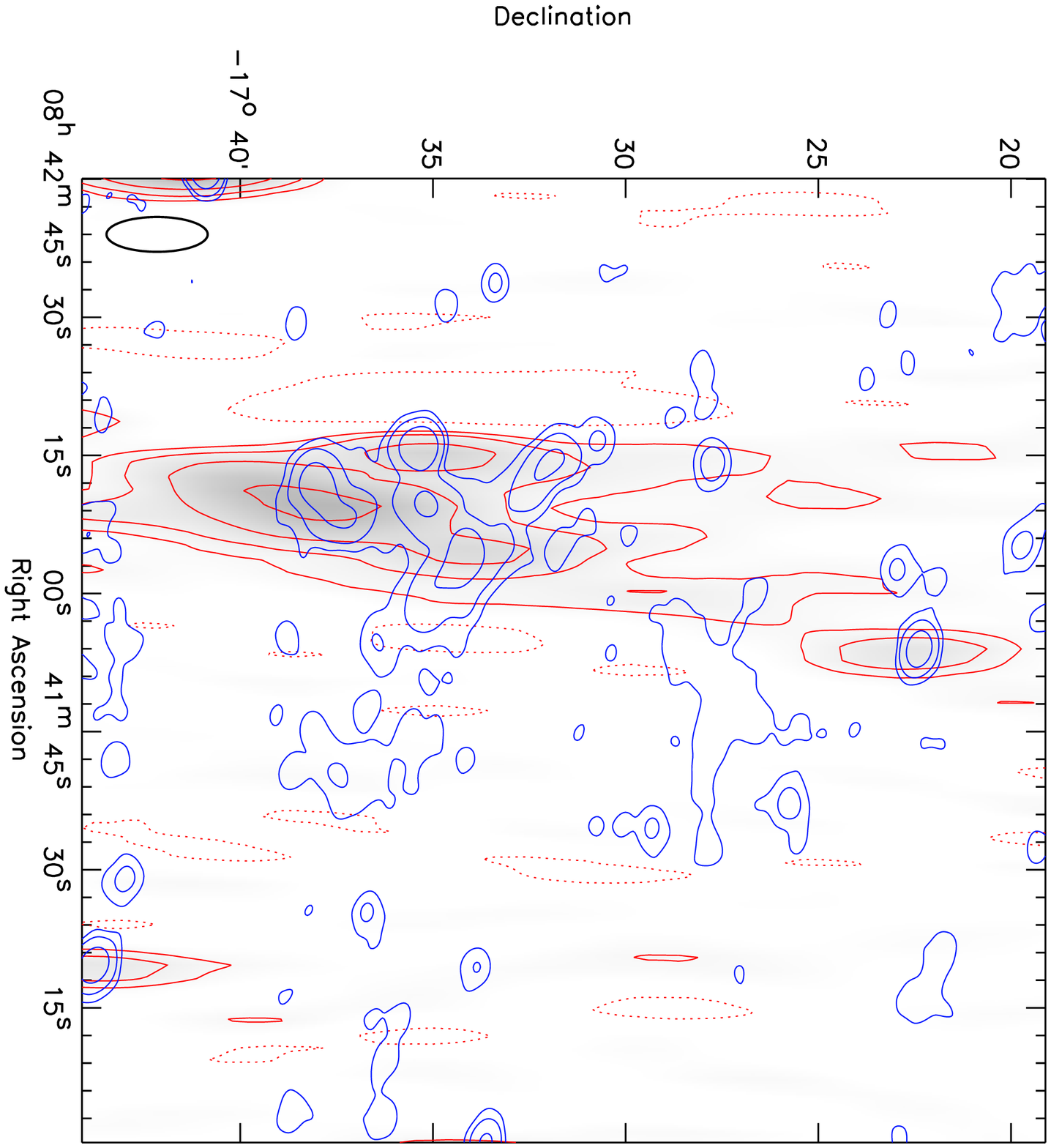}
\includegraphics[angle =90, trim =0cm 0cm 0cm 0cm,width=0.4875\textwidth]{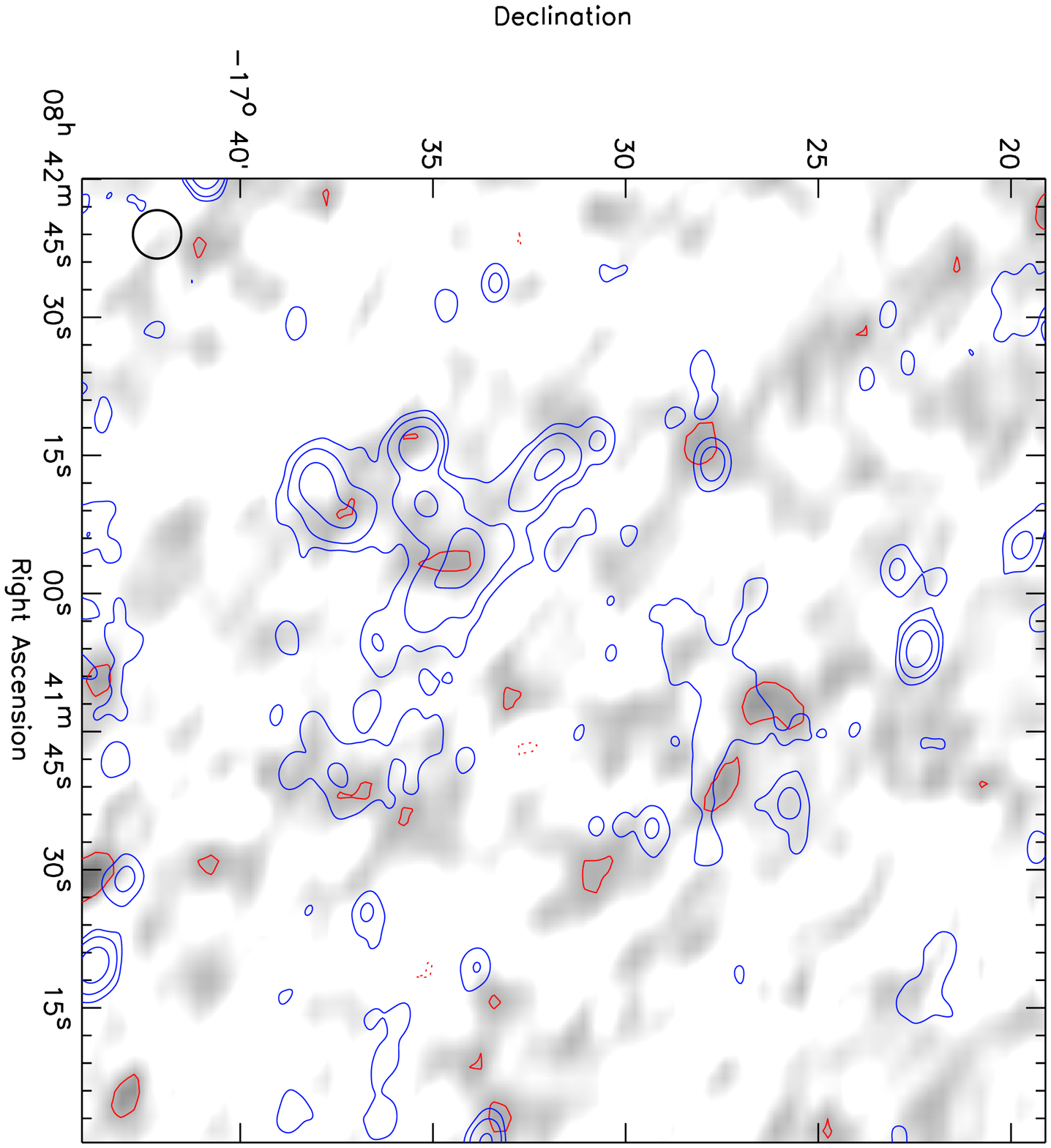}
\end{center} 
\caption{Left: 352~MHz WISH image. Red contour levels are drawn at ${[1, 2, 4, 8, \ldots]} \times 4.5$~mJy~beam$^{-1}$.  Negative  $-3\sigma_{\mathrm{rms}}$ contours are shown by the dotted lines. The image has a very elongated beam shape with a size of $157\arcsec \times 54\arcsec$. Blue contours are from the VLA DnC array image and drawn at levels of ${[1, 2, 4, 8, \ldots]} \times 0.141$~mJy~beam$^{-1}$. Right: {VLSS 74~MHz image of the cluster covering the same region as the 352~MHz WISH image. Red contours are drawn at 0.2 (solid) and $-0.3$ (dotted) Jy~beam$^{-1}$. Grayscales display the range between 0 and 0.7~Jy beam$^{-1}$ (white to black). The image has a circular beam size of 75\arcsec.}}
\label{fig:wish}
\end{figure*}

\subsection{Radio spectral index}  
The radio relic is also visible in an image from the 352~MHz Westerbork In the Southern Hemisphere (WISH) survey \citep[][see Fig.~\ref{fig:wish} left panel]{2002A&A...394...59D}. Some positive residuals are also seen in the reprocessed 74~MHz VLSS image \citep[][{Fig.~\ref{fig:wish} right panel}]{2012RaSc...47.....L}, but none of these are above the $3\sigma_{\rm{rms}}$ level. We convolved the \textit{VLA}~DnC, VLSS and WISH images to a common resolution of $157\arcsec \times 80\arcsec$. Since it is not possible to properly take into account the flux of the compact sources in the WISH and VLSS images we summed the flux over the entire cluster area (Fig.~\ref{fig:vlaimages} right panel). The radio relic flux is expected to dominate the flux contribution, since it is the brightest source at 1.4~GHz and should have a spectral index steeper than the compact  sources (mainly sources $A$ and $B$). At 1.4 GHz, the relic flux contributes 66\% to the total flux measured in the cluster area. 

\begin{figure}
\begin{center}
\includegraphics[angle =90, trim =0cm 0cm 0cm 0cm,width=0.49\textwidth]{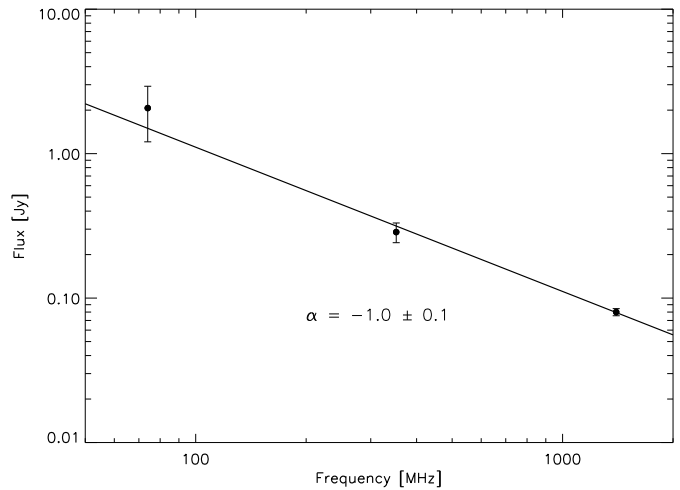}
\end{center}
\caption{Radio flux measurements of the combined flux (diffuse + compact sources) in the entire A3411 cluster region. The flux measurements were taken from the 74~MHz reprocessed VLSS survey,  the 352~MHz WISH survey, and the 1.4~GHz \textit{VLA}~DnC array image. The errors are based on the map noise, scaling with the square root of the number of beam areas, and absolute flux calibration uncertainties of 10\% for the VLSS, 15\% for WISH and 5\% for the \textit{VLA} measurement. We note that the reprocessed VLSS survey is on the \cite{2012MNRAS.423L..30S} scale so the 74~MHz flux could be a little
overestimated compared to the WISH and NVSS flux measurements \citep[see][]{2012MNRAS.423L..30S}. }
\label{fig:flux}
\end{figure}

From a power-law fit through the flux measurements at 74, 352\footnote{We took the empirical flux correction factor of $0.83$, listed in \cite{2002A&A...394...59D}, into account.}, and 1.4~GHz we find a spectral index $
\alpha=-1.0 \pm 0.1$ (see Fig.~\ref{fig:flux}), typical for bright radio relics \citep[e.g.,][]{2006AJ....131.2900C,2012A&A...546A.124V}. 

\section{Discussion}
\label{sec:discussion}

The sharp northwestern outer edge of the subcluster core, visible in the Chandra image, suggests that this subcluster fell in from the SE.  The edge itself could be a cold front, although deeper Chandra observations are needed to prove that. The presence of a large radio relic, likely tracing a shock, and the X-ray morphology imply that we currently view the merger after core passage of the subclusters. However, during a head-on binary cluster merger event two outward propagating shock waves should form along the merger axis \citep[e.g.,][]{1999ApJ...518..603R, 2001ApJ...561..621R}. Based on the X-ray morphology, a counter shock wave is expected to be traveling in front of the core visible in Fig.~\ref{fig:xray}. This region is mostly outside the FOV of the {\it Chandra} observation.
The VLA observations cover the entire region up to 2~Mpc NW of the subcluster core but no relic is found there.  
This could indicate that (i) the Mach number of this proposed counter shock is not high enough for efficient particle acceleration \citep[e.g.,][]{ 2007MNRAS.375...77H, 2012ApJ...756...97K}. Simulations indicate that for cluster mergers with mass ratios $\gtrsim3$ very little radio emission is generated by the shock wave in front of the less massive subcluster \citep{2011MNRAS.418..230V}.  From simulations it is also found that the radio emission in clusters is highly dependent on the merger state, varying on time scales of a few hundred million years \citep{2011ApJ...735...96S}. (ii) The lack of a radio relic could also imply there is no supply of pre-accelerated fossil electrons available to be re-accelerated by the shock \citep{2011MNRAS.417L...1R} because of the lack of nearby radio galaxies. 
Some of the halo emission is located NW of the well-defined subcluster core. A third possibility therefore is that we actually observe relic emission from the counter shock but it is very faint and partly mixed with halo emission. Because the NW cluster region is at the edge of the {\it Chandra} FOV and the radio emission is only visible at the $3\sigma$ level, it is currently difficult to determine the precise nature of this radio emission. Deep polarization observations could distinguish between radio halo and relic emission since relics are usually polarized at the 10--20\% level or more.

The A3411 radio relic is located at $R_{\rm{projected}}=1.3$~Mpc, a typical distance for radio relics. Few relics are found at distances beyond 2~Mpc from the cluster center \citep{2011A&A...533A..35V, 2012MNRAS.421.1868V}. If we assume that the A3411 relic is located at a distance of 2~Mpc or less from the cluster center we can constrain the line of sight of the merger geometry. The shock waves are then seen under and angle of $<50\degr$ from edge-on and the merger axis is located under an angle of $>40\degr$ along the line of sight.

The relic in A3411 has a very complex morphology, compared to some of the other clusters with single relics like Abell~521 \citep{2008A&A...486..347G}, Abell~746 \citep{2011A&A...533A..35V}, and Abell~2744 \citep{2001A&A...376..803G}.
The morphology of A3411 could have been the result of the shock front interacting with substructures or a large-scale galaxy filament from the cosmic web connected to the cluster \citep[e.g.,][]{2011ApJ...726...17P,2011MNRAS.418..230V}. Numerical simulations also show that the morphology of relics in clusters can be quite complex \citep{2011JApA...32..509H,2012MNRAS.421.1868V,2011ApJ...735...96S} and detailed observations of the single relics in for example Abell 2256, the Coma Cluster and 1RXS J0603.3+4214 also reveal more intricate morphologies \citep[e.g.,][]{1991A&A...252..528G,2006AJ....131.2900C,2012A&A...546A.124V}. The broken morphology of the relic could also be caused  by fluctuations in the magnetic field strength across the relic, which again could have been caused by varying conditions along the shock front. 

Another possibility is that the shock re-accelerates relativistic electrons trapped in several fossil radio bubbles from previous episodes of AGN activity or the relic traces the lobes of radio galaxies whose plasma has been shredded and re-accelerated. This would imply that electrons are not directly accelerated from the thermal pool \citep{2005ApJ...627..733M} but were already pre-accelerated. The fossil radio bubbles could for example be related to radio galaxy $B$ and/or $A$. Problematic is that it might be very difficult to distinguish between shock acceleration and shock re-acceleration on the basis of the spectral properties of relics.

The dynamical state of the cluster is not well known due to the lack of a sufficient number of redshift measurements. For example, it is unclear if the cD galaxy near the peak of the X-ray emission (\object{2MASX J08415287-1728046}, see Fig.~\ref{fig:xray} right panel) is associated with the well-defined subcluster core. Cluster mergers events  can decouple the baryonic matter component from the dark matter (DM) and galaxies \cite[e.g.,][]{2006ApJ...648L.109C, 2010MNRAS.406.1134S, 2012ApJ...747L..42D} due to ram pressure stripping. If this subcluster undergoes ram pressure stripping and the subcluster moves toward the NW then we would expect the cD galaxy to be located ahead of the subcluster's core, which is not the case. Therefore it is possible that the cD galaxy is associated with the primary subcluster surrounding the smaller subcluster core.
\section{Conclusions}

We have presented \textit{Chandra} and \textit{VLA} observations of the Planck ESZ cluster Abell~3411. The Chandra image reveals the cluster to be undergoing a merger event. The \textit{VLA} observations display a central 0.9~Mpc radio halo and a one-sided polarized 1.9 Mpc relic with a complex unusual morphology to the SE of the cluster. For the combined emission in the cluster area (diffuse emission + compact sources), we find a radio spectral index of $-1.0 \pm0.1$ between 74~MHz and 1.4~GHz, consistent with a bright radio relic dominating the flux contribution.
The morphology of the relic could be related to a complex-shaped shock surface due to the presence of substructures in the ICM, or reflect the presence of electrons trapped in several distinct fossil radio bubbles. This would then suggest that the synchrotron emitting electrons are re-accelerated by the shock and not accelerated directly from the thermal pool. Another possibility is that the complex morphology of the relic is caused  by fluctuations in the magnetic field strength across the relic,  for example caused by varying shock properties related to substructures. 

\label{sec:conclusions}

\acknowledgments
{\it Acknowledgments:}
{We would like to thank the anonymous referee for useful comments and Simona Giacintucci for useful discussions. 
Support for this work was provided by NASA through the Einstein Postdoctoral
grant number PF2-130104 awarded by the Chandra X-ray Center, which is
operated by the Smithsonian Astrophysical Observatory for NASA under
contract NAS8-03060. Basic research into radio astronomy at the Naval Research Laboratory is supported by 6.1 Base funds.
The scientific results reported in this article are based in part on observations made by the Chandra X-ray Observatory. 
The National Radio Astronomy Observatory is a facility of the National Science 
Foundation operated under cooperative agreement by Associated Universities, Inc.
This work is based on data products from observations made with ESO Telescopes at the La
Silla or Paranal Observatory under programme ID 070.B-0440(A).
This publication makes use of data products from the Two Micron All Sky Survey, which is a joint project of the University of Massachusetts and the Infrared Processing and Analysis Center/California Institute of Technology, funded by the National Aeronautics and Space Administration and the National Science Foundation. The Digitized Sky Surveys were produced at the Space Telescope Science Institute under U.S. Government grant NAG W-2166. The images of these surveys are based on photographic data obtained using the Oschin Schmidt Telescope on Palomar Mountain and the UK Schmidt Telescope. The plates were processed into the present compressed digital form with the permission of these institutions. The Second Palomar Observatory Sky Survey (POSS-II) was made by the California Institute of Technology with funds from the National Science Foundation, the National Geographic Society, the Sloan Foundation, the Samuel Oschin Foundation, and the Eastman Kodak Corporation.  
}



{\it Facilities:} \facility{VLA}, {CXO}.

\bibliography{ref_filaments.bib}

\end{document}